\documentclass [11pt]{article}

\usepackage{amssymb,amsmath, amsthm}
\usepackage{graphicx, color,mathrsfs}
\usepackage{latexsym}
\usepackage{amsfonts}
\usepackage{amscd}
\usepackage{txfonts}
\usepackage[all]{xy}
\usepackage{float}

 \newtheorem{thm}{Theorem}[]
 
 \newtheorem{lem}[thm]{Lemma}
 \newtheorem{prop}[thm]{Proposition}
 \theoremstyle{definition}
 
 \theoremstyle{remark}
 
\newcommand{\be}{\begin{equation}}
\newcommand{\ee}{\end{equation}}
\newcommand{\beq}{\begin{eqnarray}}
\newcommand{\eeq}{\end{eqnarray}}
\newcommand{\dis}{\displaystyle}

\newcommand{\la}{\label}

\newfont{\msbm}{msbm10 scaled\magstep1}
\newfont{\msbms}{msbm7 scaled\magstep1} 
\newif\ifintrmk

\intrmkfalse           

\begin{document}
\begin{center}
{\LARGE{Note on  the Absolutely Continuous Spectrum for the Anderson Model on Cayley Trees of Arbitrary Degree}}
\end{center}

\begin{center}
{\Large {Florina Halasan}} \footnote{Email address: florina.halasan@gmail.com\\  The author was supported by an NSERC/MITACS Canadian Scholarship. This paper is based on part of the author's doctoral thesis.\\ It is a pleasure to thank Dr. R. Froese for all his guidance and support.}
\end{center}

\bigskip \bigskip \bigskip
  
\begin{abstract} We provide a simplified version of the geometric method given by Froese, Hasler and Spitzer in~\cite{FHS-2} and use it to prove the existence of absolutely continuous spectrum for a Cayley tree of arbitrary degree $k$.
\end{abstract}



\section{Introduction}

One of the most important open problems in the field of random Schr\"odinger operators is to prove the existence of absolutely continuous spectrum for weak disorder in the Anderson model \cite {And} in three and higher dimensions. The first result in this direction is Abel Klein's, for random Schr\"odinger operators acting on a tree, or Cayley tree, or Bethe lattice, of any constant degree larger than $2$. Klein \cite{K} proves that for weak disorder, almost all potentials will produce absolutely continuous spectrum. This means that there must be many potentials on a tree for which the corresponding Schr\"odinger operator has absolutely continuous spectrum without there being an obvious reason, such as periodicity or decrease at infinity. Later on, different other proofs were given to the same result (see \cite{FHS-2} and \cite{ASW}). This paper simplifies the geometric method in \cite{FHS-2}. The simplifications make possible the generalization from a Bethe lattice of degree $3$ to one of any degree $M+1$, with $M \geq 2$. 

\section{The Model and the Results}

A Bethe lattice (or Cayley tree), $\mathbb{B}$,  is a connected infinite graph with no closed loops and a fixed degree (number of nearest neighbors) at each vertex (site or point), $x$. The distance between two sites $x$ and $y$ will be denoted by $d(x,y)$ and is equal to the length of the shortest (only) path connecting $x$ and $y$.

The Anderson model on the Bethe lattice is given by the random Hamiltonian 
$$ 
H\;\;=\;\;\Delta\, + k\,q 
$$ 
on the Hilbert space $\dis \ell^2(\mathbb{B})=\{\varphi:\mathbb{B} \rightarrow \mathbb{C}\,;\,\sum_{x\in \mathbb{B}} \left| \varphi(x)\right|^2<\infty \}$. The (centered) Laplacian $\Delta$ is defined by
\begin{equation*}
(\Delta \varphi)(x)=\sum_{y:\,d(x,y)=1} \varphi(y)
\end{equation*}
and has spectrum $ \sigma(\Delta) = [-2\sqrt{ M},2\sqrt{M}]\,$. The operator $q$ is a random potential, with $q(x)$, $x \in \mathbb{B}$, being independent, identically distributed real random variables with common probability distribution $\nu$. We assume the $2(1+p)$ moment, $\dis \int |q|^{2(1+p)}d\nu$, is finite for some $p>0$. The coupling constant $k$ measures the disorder.

As mentioned above, the existence of purely absolutely continuous spectrum for the Anderson model on the Bethe lattice was first proved, in a different manner, by Klein in $1998$. Given any closed interval $E$ contained in the interior of the spectrum of $\Delta$ on the Bethe lattice, he proved that for small disorder, $H$ has purely absolutely continuous spectrum in some interval $E$ with probability one, and its integrated density of states is continuously differentiable on the interval (he only needed a finite second moment, whereas we have a finite 2(1+p) moment in our model). We prove a similar result in this chapter. A key point is the definition of a weight function appearing in the proofs. This definition is motivated by hyperbolic geometry. 
\smallskip

\begin{thm} \la{T2.1} For any $E$, with $0 < E <2 \sqrt{M}$ and $H$ defined above, there exists $k(E) > 0$ such that for all $0<|k|<k(E)$ the spectrum of $H$ is purely absolutely continuous in $[-E,E]$
with probability one, i.e., we have almost surely
$$
\sigma_{\rm ac} \cap [-E,E] = [-E, E] \ , \ \ \sigma_{\rm pp} \cap
[-E,E] = \emptyset \ , \ \ \sigma_{\rm sc} \cap [ -E, E] =
\emptyset \; .
$$
\end{thm}

\noindent Let $\mathbb{H} = \{z\in \mathbb{C}: {\rm Im}(z) > 0\}$ denote the complex upper half plane. For convenience, we fix an arbitrary site in $\mathbb{B}$ to be the origin and denote it by $0$. For each $x \in \mathbb{B}$ we have at most one neighbour towards the root and two or more in what we refer to as the forward direction. We say that $y \in \mathbb{B}$ is in the future of $x \in \mathbb{B}$ if the path connecting $y$ and the root runs through $x$.  Let $x \in \mathbb{B}$ be an arbitrary vertex, the subtree consisting of all the vertices in the future of $x$, with $x$ regarded as its root, is denoted by $\mathbb{B}^{x}$. We will write $H^{x}$ for $H$ when
restricted to $\mathbb{B}^{x}$ and set
$ \dis
G^{x}(\lambda)=\langle {\delta_x,(H^{x} -\lambda)^{-1}\delta_x}
\rangle 
$ the Green function for the truncated graph. $G^x$ is called the forward Green function.

\begin{prop} For any $\lambda \in \mathbb{H}$ we have
\begin{equation} G(\lambda)=\left\langle {\delta_0,(H -\lambda)^{-1}\delta_0}
\right\rangle= -\,\left(\sum_{x:\,d(x,0)=1}
 G^{x}(\lambda)+\lambda - k\,q(0) \right)^{-1}\;  \la{2.1}
\end{equation} and, for any site $x \in \mathbb{B}$, 
\begin{equation}
 G^{x} (\lambda)= -\left(\sum_{y:\,d(y,x)=1,\,y\in \mathbb{B}^x} G^{y}\,  (\lambda)+\lambda - k\,q(x)
\right)^{-1}\;.  \la{2.2} \end{equation} 
\end{prop}

\proof  We will prove (\ref{2.1}); (\ref{2.2}) is proven in exactly
the same way.  Let us write $\dis H =\tilde{H} + \Gamma$, where 
\begin{eqnarray*} 
\tilde{H}&=& k\,q(0) \oplus \left(\bigoplus_{x:\,d(x,0)=1} H^{x}\right) 
\end{eqnarray*} 
is the direct sum corresponding to the decomposition $\mathbb{B}=\{0\}\cup
\left(\bigcup\limits_{x:\,d(x,0)=1}{\mathbb{B}}^{x}\right)$.  The
operator $\Gamma$ has matrix elements $\left\langle {\delta_x,\Gamma \delta_0} \right\rangle=\left\langle {\delta_0,\Gamma\delta_x} \right\rangle= 1$
if $d(x,0)=1$, with all other matrix elements being $0$.  The
resolvent identity gives \begin{eqnarray*} (\tilde{H} -\lambda)^{-1}&=& (H
-\lambda)^{-1}+ (\tilde{H} -\lambda)^{-1} \Gamma\, (H -\lambda)^{-1}\;. 
\end{eqnarray*}
Also, 
\begin{eqnarray*} 
(\tilde{H} -\lambda)^{-1}&=& (k\,q(0) -\lambda)^{-1} \oplus \left(\bigoplus_{x:\,d(x,0)=1}( H^{x} -\lambda)^{-1}\right)\,.
\end{eqnarray*}  
Thus
\begin{eqnarray*} 
\left\langle {\delta_0,(\tilde{H} -\lambda)^{-1}\delta_0}
\right\rangle&=&\left\langle {\delta_0,(H -\lambda)^{-1}\delta_0}
\right\rangle + \left\langle {\delta_0,(\tilde{H}
-\lambda)^{-1}\Gamma\,(H-\lambda)^{-1}\delta_0} \right\rangle\;. 
\end{eqnarray*} 
Hence 
\begin{eqnarray}
G\,(\lambda)&=&(q(0)- \lambda)^{-1} - (k\,q(0)-
\lambda)^{-1}\sum_{x:\,d(x,0)=1}
 \left\langle {\delta_x,(H -\lambda)^{-1}\delta_0}
\right\rangle\;,  \la{2.3}
\end{eqnarray} 
The resolvent formula also implies that for each $x$ with $d(x,0)=1$, 
\begin{equation}
 \left\langle {\delta_x,(H -\lambda)^{-1}\delta_0}
\right\rangle= -  G^{x}\,  (\lambda)\, G\, (\lambda) \;. \la{2.4}
\end{equation}
 (\ref{2.2}) follows from  (\ref{2.3}) and  (\ref{2.4}).  \qed

The recursion relation for $G^{x}(\lambda)$ that we just proved leads us to the following transformation
$$
\phi:\mathbb{H}^{M}\times \mathbb{R} \times \mathbb{H} \to \mathbb{H}
$$
defined by 
\begin{equation} \la{2.5}
\phi(z_1,...z_M,q,\lambda) = \frac{-1}{z_1+...+z_M+\lambda-q}\; . 
\end{equation} 
It is easy to see the equivalence between (\ref{2.1}) and (\ref{2.5}). Let $q \equiv 0$.
If ${\rm Im}( \lambda) > 0$, the transformation $z\mapsto\phi(z,...,z,0,\lambda)$ has a
unique fixed point, $z_\lambda$,  in the upper half plane, i.e. ${\rm Im}( z_\lambda) > 0$ (for details see Proposition $2.1$, in~\cite{FHS-1}). Explicitly,
$$
z_\lambda = \frac{-\lambda}{2M} +
\frac{1}{M}\sqrt{(\lambda/2)^2-M} \; ,
$$
where we will always make the choice ${\rm Im} \sqrt{ \, \cdot \,} \geq 0$ (and $\sqrt{ a} > 0$ for $a>0$). This fixed point as a function of $\lambda \in \mathbb{H}$ extends continuously onto the real axis. This extension yields, for $\rm{Im}(\lambda) = 0$ and $|\lambda|< 2 \sqrt{M}$, the fixed point
$$
z_\lambda = -\frac{\lambda}{2M} + \frac{i}{2M} \sqrt{4M -\lambda^2} \;,
$$
lying on an arc of the circle $|z|=1/\sqrt{M}$. When $\rm{Im}
(\lambda) = 0$ and $|\lambda|\leq E < 2 \sqrt{M}$, the arc is
strictly contained in the upper half plane. Thus, when $\lambda$
lies in the strip
$$
R(E,\epsilon) = \{z\in \mathbb{H} : {\rm Re}(z) \in [-E,E], 0<{\rm
Im}(z) \leq \epsilon \}
$$
with $0<E<2\sqrt{M}$ and $\epsilon$ sufficiently small, ${\rm
Im}(z_\lambda)$ is bounded below and $|z_\lambda|$ is bounded
above by a positive constant.

In order to prove that the spectral measures are absolutely continuous we need to establish bounds for $\mathbb{E}(|G^x(\lambda)|^{1+p})$. Since $z_\lambda$ equals $G^x(\lambda)$ for the case $ q \equiv 0$ and any $x \in \mathbb{B}$, in order to prove the desired bounds we will use the weight function $\mathrm{w}(z)$ defined by
\begin{equation} 
\mathrm{w}(z) = 2 \,(\cosh({\rm
dist}_{\mathbb{H}}(z,z_\lambda))-1) = {{|z-z_\lambda|^2}\over{{\rm
Im}(z)}{\rm Im}(z_\lambda)} \; . 
\end{equation} 
Up to constants, $\mathrm{w}(z)$ is the hyperbolic cosine of the hyperbolic distance from $z$ to $z_\lambda$, provided $\lambda\in R(E,\epsilon)$ with $0<E<2\sqrt{M}$ and $\epsilon$ sufficiently small. This notation suppresses the $\lambda$ dependence. In essence, we are looking at the hyperbolic cosine of the distance between $G^x(\lambda)$ for the free Laplacian and the one for the perturbed one, $H$. The goal is to prove that this quantity, which blows up on the boundary, stays mostly finite. 

To prove a bound for $\mathbb{E}(\mathrm{w}^{1+p}(G^x(\lambda)))$ we will need to use (\ref{2.5}), more than once, to express the forward Green function as a function of the forward Green functions at future nodes. As a result, the study of the following quantity becomes needed:
\begin{eqnarray*}
\mu_{3,p}(z_1\ldots z_{2M-1},q_1,q_2, \lambda) = \sum_{\sigma}
{{\mathrm{w}^{1+p}(\phi(\phi(z_{\sigma_1}\ldots z_{\sigma_M},
q_1, \lambda),z_{\sigma_{M+1}}\ldots z_{\sigma_{2M-1}}
q_2,\lambda))}\over{ \mathrm{w}^{1+p}(z_1) +\ldots+ {\rm \mathrm{w}}^{1+p}(z_{2M-1}) }}
\end{eqnarray*}
where $\sigma$ are all cyclic permutations. We will state here the needed lemmas, but we will give the proofs later.

\begin{lem} \la{L2.1} For any $E$, $0<E<2\sqrt{M}$ and any $0<p<1$, there
exist positive constants $\epsilon$, $\eta_1$, $\epsilon_0$ and a
compact set ${\cal M} \in \mathbb{H}^{2M-1}$ such that
\begin{eqnarray}
  \mu_{3,p}|_{{\cal M} ^c \times [-\eta_1, \eta_1]^2 \times R(E,\epsilon_0 )} &\leq& 1- \epsilon .
\end{eqnarray}
Here ${\cal M} ^c$ denotes the complement $\mathbb{H}^{2M-1} \setminus
{\cal M}$.
\end{lem}

\begin{lem} \la{L2.2} For any $E$, $0<E<2\sqrt{M}$ and any $0<p<1$, there
exist positive constants $\epsilon_0$, $C$ and a compact set
${\cal M} \in \mathbb{H}^{2M-1}$ such that
\begin{eqnarray}
  \mu_{3,p}|_{{\cal M}^c \times \mathbb{R}^2 \times R(E,\epsilon_0 )} &\leq& C ( 1+ \sum_{i=1}^2 |q_i|^{2(1+p)}) .
\end{eqnarray}
Similarly, if we define
\begin{eqnarray*}
  \mu'_{3,p}(z_1,\ldots,z_{M+1}) &=& \frac{\mathrm{w}(-(\sum\limits_{i=1}^{M+1}z_i+\lambda-q)^{-1})^{1+p}}{\mathrm{w}(z_1)^{1+p} + \ldots+ \mathrm{w}(z_{M+1})^{1+p} }\,, 
\end{eqnarray*}
then
\begin{eqnarray*}
   \mu'_{3,p}|_{{\cal M}^c \times \mathbb{R}^2 \times R(E,\epsilon_0 )}&\leq& C ( 1+ |q|^{2(1+p)})\,.
\end{eqnarray*}
\end{lem}
\newpage

\begin{thm} \la{T2.4} Let $x$ be a nearest neighbour of $0$. For any $E$,
$0<E<2\sqrt{M}$ and all $0<p<1$, there exists $k(E) > 0$ such that for all $0<|k|<k(E)$ we have
$$
\sup_{\lambda \in R(E,\epsilon)} \mathbb{E} \left(  \mathrm{w}^{1+p}(G^{x}(\lambda)) \right)
 < \infty \; .
$$
\end{thm} 
\proof In order to prove that the above quantity is bounded we need a couple of preparatory steps. 

Let $\eta_1$ and $p$ be given by Lemma \ref{L2.1},
and choose $\epsilon_0$ and $\cal M$ that work in both Lemma
\ref{L2.1} and Lemma \ref{L2.2}. For $(z_1,\ldots,z_{2M-1})\in
{\cal M}^c$, we estimate
\begin{align*}
&\int_{\mathbb{R}^2}\mu_{3,p}(z_1,\ldots,z_{2M-1},k\,q_1,k\,q_2,\lambda)
d\nu(q_1) d\nu(q_2) \\
&\leq (1-\epsilon)\int_{\left[\frac{-\eta_1}{k},\frac{\eta_1}{k}\right]^2}d\nu(q_1) d\nu(q_2)
+C\int_{\mathbb{R}^2\backslash \left[\frac{-\eta_1}{k},\frac{\eta_1}{k}\right]^2}
(1+\sum_{i=1}^2|k\,q_i|^{2(1+p)})\,d\nu(q_1) d\nu(q_2) \\
&\leq (1-\epsilon) + C\int_{\mathbb{R}^2\backslash \left[\frac{-\eta_1}{k},\frac{\eta_1}{k}\right]^2}\,d\nu(q_1) d\nu(q_2) + 2C |k|^{2(1+p)}M_{2(1+p)} \leq 1-\epsilon/2 
\end{align*}
provided $k$ is sufficiently small. Here $M_{2(1+p)}$ denotes
the moment $\int |q|^{2(1+p)}\,d\nu(q)$.

The probability distributions for $G$ and $G^x$ on the hyperbolic plane are defined by $\dis \rho_G(A) = {\rm Prob} \{ G(\lambda) \in A \}$ and $\dis \rho(A) = {\rm Prob} \{ G^{x}(\lambda) \in A \}$. This implies
\begin{align*}
&\rho(A) = {\rm Prob}\{\phi(z_1\ldots z_M,k\,q,\lambda)\in A \} =
{\rm Prob}\{(z_1 \ldots z_M,k\,q,\lambda)\in \phi^{-1}(A) \}\\
&=\int\limits_{\phi^{-1}(A)} d\rho(z_1)\ldots d\rho(z_M)\,d\nu(q) 
=\int\limits_{\mathbb{H}^M\times\mathbb{R}} \chi_A
(\phi(z_1\ldots z_M,k\,q,\lambda))\,d\rho(z_1) \ldots d\rho(z_M)\,d\nu(q)
\end{align*}
which gives us that for any bounded continuous function $\mathrm{w}(z)$,
\begin{align*}
&\int_{\mathbb{H}}\mathrm{w}(z) d\rho(z) = \int_{\mathbb{H}^M\times\mathbb{R}}
\mathrm{w}(\phi(z_1,\ldots,z_M,k\,q,\lambda))\,d\rho(z_1)\,\ldots\,d\rho(z_M)\,d\nu(q).
\end{align*}
Now we have all the ingredients needed to prove our theorem. Using the previous relation twice, for $\lambda \in R(E,\epsilon_0)$, we obtain:
\newpage
\begin{align*}
&\mathbb{E}\left(\mathrm{w}^{1+p}(G^x(\lambda)) \right) = \int\limits_{\mathbb{H}} \mathrm{w}^{1+p}(z)\, d \rho(z) \\
&= \int\limits_{\mathbb{H}^M\times\mathbb{R}} \mathrm{w}^{1+p}(\phi(z_1\ldots z_M,k\,q_1,\lambda))\,
d\rho(z_1) \ldots d\rho(z_M)\, d\nu(q_1) \\
&= \int\limits_{\mathbb{H}^M\times\mathbb{R}^2} \mathrm{w}^{1+p}(\phi(\phi(z_1\ldots z_M,k\,q_1,\lambda),z_{M+1} \ldots z_{2M-1},k\,q_2,\lambda))\,\\
&\hspace{8.5cm} d\rho(z_1)\ldots d\rho(z_{2M-1} )\, d\nu(q_1) d\nu(q_2) \\
&=\int\limits_{\mathbb{H}^M\times\mathbb{R}^2} \frac{1}{2M-1} \sum_\sigma {\rm \mathrm{w}}^{1+p}\big(\phi(\phi(z_{\sigma_1} \ldots z_{\sigma_M},k\,q_{1},\lambda),z_{\sigma_{M+1}}\ldots z_{\sigma_{2M-1}}, k\,q_{2}, \lambda))\big)\,\\
&\hspace{8.5cm} d\rho(z_1) \ldots d\rho(z_{2M-1})\,
d\nu(q_1) d\nu(q_2) 
\end{align*}
\begin{align*}
&= \frac{1}{2M-1} \int_{{\cal M}^c}\left(\int_{\mathbb{R}^2}
\mu_{3,p}(z_1 \ldots z_{2M-1},k\,q_1,k\,q_2,\lambda)\,d\nu(q_1)d\nu(q_2)\right)\\  
&\,\,\,\,\,\,\,\,\,\,\,\,\,\,\,\,\,\,\,\,\,\,\,\,\,\,\,\,\,\,\,\,\,\,\,\,\,\,\,\,\,\,\,\,\,\,\,\,\,\,\,\,\,\,\,\, \quad \times \big( \mathrm{w}^{1+p}(z_1) + \ldots + \mathrm{w}^{1+p}(z_{2M-1}) \big)\,d\rho(z_1) \ldots  d\rho(z_{2M-1}) + C \\
&\leq ( 1 - \epsilon/2) \int_{\mathbb{H}} \mathrm{w}^{1+p}(z)\, d \rho(z) + C = (1-\epsilon/2)\,\mathbb{E}\left(\mathrm{w}^{1+p}(G^x(\lambda) \right)+C \, .
\end{align*}
where $C$ is some finite constant, only depending on the choice of $\cal M$.

\noindent{\it Note}: We used the fact that $$ \int_{\mathbb{H}} \mathrm{w}^{1+p}(z) d \rho(z) = \frac{1}{2M-1} \int_{\mathbb{H}^{2M-1}} \left(\mathrm{w}^{1+p}(z_1)+\ldots+\mathrm{w}^{1+p}(z_{2M-1})\right) d \rho(z_1)\ldots d\rho(z_{2M-1})$$

 \noindent This implies that for all $\lambda \in R(E,\epsilon_0)$,
$$
\mathbb{E} \left( \mathrm{w}^{1+p}( G^{x}(\lambda))  \right) \leq \frac{2C}
{\epsilon}  \; .
$$ \qed

\begin{thm} Let $x \in \mathbb{B}$. Under the hypotheses of Theorem
\ref{T2.4},
$$
\sup_{\lambda \in R(E,\epsilon) } \mathbb{E}  \left( \left| \langle
\delta_x , (H - \lambda)^{-1}  \delta_x \rangle \right|^{1+p} \right) < \infty \;
$$
for some $\epsilon > 0$. 
\end{thm}

\proof It is an immediate consequence of Theorem \ref{T2.4} and the following inequality:
\begin{equation} \la{ineq:abs}
|z| \leq 4 {\rm Im}(s){{|z-s|^2}\over{{\rm Im}(z){\rm Im}(s)}} + 2 |s| \; .
\end{equation}
The inequality clearly holds for $|z| \leq 2 |s|$. In the complementary case, we have $|z| > 2 |s|$ and thus $|z-s| \geq ||z|-|s|| \geq |s|$, implying $\dis |z|{\rm Im}(z) \leq |z|^2 \leq 2 |z-s|^2 + 2 |s|^2 \leq 4 |z -s|^2$.  This proves (\ref{ineq:abs}).

Using (\ref{ineq:abs}) with $s=z_\lambda$ yields that for
$\lambda\in R(E,\epsilon)$, $\dis |z| \le 4 \mathrm{w}(z) + C$, where $C$ depends only on $E$ and $\epsilon$. 

To finish the proof we need to transfer the estimate from $\rho$ to $\rho_G$ and therefore prove the inequality for $x=0$. By symmetry it extends to any vertex $x \in \mathbb{B}$. In the proof of the following estimate we need the elementary fact that for $z_1$\ldots $z_{M+1} \in \mathcal{M}$, 
$\dis \mathrm{w}^{1+p}\left(\left(\sum\limits_{i=1}^{M+1} z_i+\lambda - q\right)^{-1}\right) \leq C \left(1+|q|^{2(1+p)}\right)$. Let $R$ denote $R(E,\epsilon)$, then

\begin{align*}
& \sup\limits_{\lambda \in R} \mathbb{E} \left( \left| \left\langle \delta_0 ,
(H-\lambda )^{-1}  \delta_0 \right\rangle \right|^{1+p} \right)= \sup\limits_{\lambda \in R} \int\limits_\mathbb{H} |z|^{1+p} \,d\rho_G(z)\\ 
 &\leq  C_1 \, \sup\limits_{\lambda \in R} \int_\mathbb{H} \mathrm{w}^{1+p}(z)\,d\rho_G(z) + C_2 \\ 
 &= C_1 \, \sup\limits_{\lambda \in R} \int\limits_{\mathbb{H}^{M+1}\times\mathbb{R}} \mathrm{w}^{1+p}\left(\left(\sum\limits_{i=1}^{M+1} z_i+\lambda - k\,q\right)^{-1}\right)\, d\rho(z_1) \ldots d\rho(z_{M+1}) \,d\nu(q) + C_2 \\ 
 &\le C_1 \, \sup\limits_{\lambda \in R} \int\limits_{{\cal M}^c\times\mathbb{R}}
\mu'_{3,p}(z_1,\ldots,z_{M+1},k\,q,\lambda) \times(\mathrm{w}^{1+p}(z_1)+\ldots+\mathrm{w}^{1+p}(z_{M+1}))\\
&\hspace{8cm} d\rho(z_1) \ldots d\rho(z_{M+1}) \,d\nu(q) + C'_2 \\ 
&\le C \int\limits_{\mathbb{H}\times\mathbb{R}}(1+|k\,q|^{2(1+p)})\mathrm{w}^{1+p}(z) \,d\rho(z)\,d\nu(q) +C_2 \le C\int\limits_\mathbb{H} \mathrm{w}^{1+p}(z) \,d\rho(z) + C_3\\
& = C\, \mathbb{E} \left( \mathrm{w}^{1+p}( G^{x}(\lambda))  \right) +C_3 \leq C_4\,,
\end{align*} 
where $C$, $C_1$, $C_2$,$C_3$ and $C_4$ are positive constants. \qed

\noindent As it was proven in~\cite{H} (or in the next chapter), this theorem implies the main result of this chapter:
\noindent {\bf Theorem \ref{T2.1}.} {\it For any $E$, with $0 < E <2 \sqrt{M}$, there exists $k(E) > 0$ such that for all $0<|k|<k(E)$ the spectrum of $H$ is purely absolutely continuous in $[-E,E]$
with probability one, i.e., we have almost surely
$$
\sigma_{\rm ac} \cap [-E,E] = [-E, E] \ , \ \ \sigma_{\rm pp} \cap
[-E,E] = \emptyset \ , \ \ \sigma_{\rm sc} \cap [ -E, E] =
\emptyset \; .
$$}

\section{Analysis of $\mu_2$ and Proofs of Lemmas}

For the proofs of our technical lemmas we need to analyse a quantity, $\mu_2$, which will prove to play a significant role in the expression for $\mu_{3,p}$.  We define $\mu_2$ by
\begin{equation*}
\mu_2(z_1\ldots z_M,q,\lambda) = {{M\,\mathrm{w}(\phi(z_1 \ldots z_M,q,\lambda))}\over{\mathrm{w}(z_1)+\ldots+\mathrm{w}(z_M)}}
\end{equation*}
as a function from
$\mathbb{H}^M\backslash\{(z_\lambda,\ldots,z_\lambda)\}\times \mathbb{R}\times
R\rightarrow \mathbb{R}$. In this section $R=R(E,\epsilon)$, for some
$0<E<2\sqrt{M}$ and $\epsilon>0$.

\begin{prop} \la{Prop2} For all $z_1,\ldots,z_M\in
\mathbb{H}^M\backslash\{(z_\lambda,\ldots,z_\lambda)\}$ and $\lambda\in
R$,
$$
\mu_2(z_1,\ldots,z_M,0,\lambda) < 1\,.
$$
\end{prop}

\proof For $z,s\in\mathbb{H}$ set
$$
{\rm c}(s,z) = 2({\rm cosh}({\rm dist}_{\mathbb{H}}(s,z))-1) =
{{|s-z|^2}\over{{\rm Im}(s){\rm Im}(z)}}\,.
$$
Note that $z\mapsto {\rm c}(s,z)$ is strictly convex. This can be
seen for example by noting that its Hessian has strictly positive
eigenvalues. Also, for $s=z_\lambda$, $\mathrm{c}(z_\lambda,z) = \mathrm{w}(z)$. The transformation $\phi'(z)= - 1/(z+\lambda) $ is a hyperbolic contraction (see ~\cite{FHS-1}, Proposition 2.1) and since  $\phi'(z_1+\ldots+z_M)=\phi(z_1 \ldots z_M,0,\lambda)$ we have $\phi'(M z_\lambda)= z_\lambda$.  This implies
\begin{align*}
 & {\rm dist}_{\mathbb{H}}(\phi'(Mz_{\lambda}), \phi'(z_1+\ldots+z_M)) < {\rm dist}_{\mathbb{H}}(Mz_{\lambda},z_1+\ldots+z_M) \Leftrightarrow\\
  &{\rm cosh}({\rm dist}_{\mathbb{H}}(\phi'(Mz_{\lambda}), \phi'(z_1+\ldots+z_M))) < {\rm cosh}({\rm dist}_{\mathbb{H}}(Mz_{\lambda},z_1+\ldots+z_M)) \Leftrightarrow  \\
& c(z_{\lambda}, \phi(z_1,\ldots,z_M,0,\lambda))< c(Mz_{\lambda}, z_1+\ldots+z_M) = c \left( z_{\lambda}, \frac{(z_1+\ldots+z_M)}{M}\right)   \\
  &\,\,\,\,\,\,\,\,\,\,\,\,\,\,\,\,\,\,\,\,\,\,\,\,\,\,\,\,\,\,\,\,\,\,\,\,\,\,\,\,\,\,\,\,\leq{{1}\over{M}} \sum_{i=1}^M c(z_{\lambda},z_i) \,,
  \end{align*}
  hence
\begin{align*}  
 & \frac{Mc(z_\lambda, \phi(z_1,\ldots,z_M,0,\lambda))}{\sum\limits_{i=1}^M c(z_{\lambda},z_i)}<
  1 
\end{align*}
Also, from Proposition $2.1$~\cite{FHS-1}, if ${\rm Im}(\lambda)=0$ then
$\phi'$ is a hyperbolic isometry. Therefore
\begin{eqnarray*}
  c(\phi'(M z_{\lambda}),\phi'(z_1+\ldots+z_M)) &=& c(M z_{\lambda},z_1+\ldots+z_M) \\
 &=&  c\left(z_{\lambda}, {{z_1+\ldots+z_M}\over{M}}\right) \leq
 {{1}\over{M}} \sum_{i=1}^M c(z_{\lambda},z_i)
\end{eqnarray*}
If ${\rm Im}(\lambda)=0$, then $\mu_2(z,\ldots,z,0,\lambda)=1$. If
${\rm Im}(\lambda)>0$, since $\phi'$ is a hyperbolic contraction,
$\mu_2(z,\ldots,z,0,\lambda)=1$ iff
$z_1=\ldots=z_M=z_{\lambda}$.\qed

\smallskip

Since in our lemmas we will use a compactification argument, we need to understand the behavior of $\mu_2(z_1,\ldots,z_M,q,\lambda)$ as $z_1$,\ldots,$z_M$ approach the boundary of $\mathbb{H}$ and $\lambda$ approaches the real axis. Thus, it is natural to introduce the compactification
$\overline{\mathbb{H}}^M\times \mathbb{R}\times \overline{R}$. Here $ \overline{R}$ denotes the closure and $\overline{\mathbb{H}}$ is the compactification of $\mathbb{H}$ obtained by adjoining
the boundary at infinity. (The word compactification is not quite accurate here because of the factor $\mathbb{R}$, but we will use the term nevertheless.)

The boundary at infinity is defined as follows.  We cover the upper half plane model of the hyperbolic plane $\mathbb{H}$ with the atlas $\mathcal{A}=\{(U_i, \psi_i)_{i=1,2}\}$. We have $U_1=\{z\in \mathbb{C} : {\rm Im}(z) > 0, |z| < C\}$, $\psi_1(z)=z$, $U_2 =\{z\in \mathbb{C} : {\rm
Im}(z) > 0, |z| > C\}$ and $\psi_2(z)=-1/z=w$. The boundary at infinity consists of the sets $\{{\rm Im}(z)=0\}$ and $\{{\rm Im}(w)=0\}$ in the respective charts. The compactification $\overline{\mathbb{H}}$ is the upper half plane with the boundary at infinity adjoined. We will use $i\infty$ to denote the point where $w=0$.

With this convention, $\mu_2$ is defined in the interior of the compactification $\overline{\mathbb{H}}^M\times \mathbb{R}\times \overline{R}$ and we want to know how it behaves near the boundary. It turns out that in the coordinates introduced above, $\mu_2$ is a rational function. For the majority of points on the boundary the denominator does not vanish in the limit and $\mu_2$ has a continuous extension. There are, however, points where both numerator and denominator vanish and at these singular points the limiting value of $\mu_2$ depends on the direction of approach. By blowing up the singular points, it would be possible to
define a compactification to which $\mu_2$ extends continuously. However, this is more than we need for our analysis. We will do a partial resolution of the singularities of $\mu_2$ and then extend $\mu_2$ to an upper semi-continuous function on the resulting compactification.

The reciprocal of the function $\mathrm{w}(z)$, $\dis \chi(z) = {{1}\over{w(z)}} = {{{\rm Im}(z){\rm Im}(z_{\lambda})}\over{|z-z_{\lambda}|^2}}$ is a boundary defining function for $\mathbb{H}$. This means that in
each of the two charts above, $\chi$ is positive near infinity and vanishes exactly to first order on the boundary at infinity. Further more, we can express $\mu_2$ as follows:
\begin{eqnarray*}
\mu_2(z_1 \ldots z_M,q,\lambda)&=&\frac{M}{\chi(\phi(z_1 \ldots z_M,q,\lambda))[\frac{1}{\chi(z_1)}+\ldots+\frac{1}{\chi(z_M)}]} 
\end{eqnarray*}
or
\begin{align*}
& \mu_2(z_1\ldots z_M,q,\lambda)= \frac{M\chi(z_1)\ldots\chi(z_M)}{\chi(\phi(z_1\ldots z_M,q,\lambda))[\chi(z_1)\ldots\chi(z_{M-1})+\ldots+\chi(z_2)\ldots+\chi(z_M)]}
\end{align*}
Since
$$
\chi(\phi(z_1\ldots z_M,q,\lambda))=\frac{{\rm
Im}(\phi(z_1\ldots z_M,q,\lambda))}{|z_{\lambda}-\phi(z_1 \ldots z_M,q,\lambda)|^2}=\frac{{\rm
Im}(z_1+\ldots+z_M+\lambda)}{|z_{\lambda}(z_1+\ldots+z_M)+\lambda
z_{\lambda}-q z_{\lambda}+1|^2}
$$
we obtain
\begin{equation}
\mu_2(z_1 \ldots z_M,q,\lambda)= \frac{M \prod\limits_{i=1}^M
\chi(z_i) |z_{\lambda}\sum\limits_{i=1}^M z_i+\lambda
z_{\lambda}-q z_{\lambda}+1|^2}{[\sum\limits_{j=1}^M
\prod\limits_{\substack {i=1\\i\neq j}}^M \chi(z_i)]
[\sum\limits_{i=1}^M\chi(z_i) |z_i-z_\lambda |^2 + {\rm
Im}(\lambda)]} \la{Mu2gen}
\end{equation}

We will now describe our compactification of $\mathbb{H}^M\times
\mathbb{R}\times R$. Start with $\overline{\mathbb{H}}^M\times \mathbb{R}\times \overline{R}$. Our blow-up consists of writing $\chi(z_1),\ldots,\chi(z_M)$ in
polar co-ordinates. Thus we introduce new variables $r_1$ and
$\beta_i$ and impose the equations $\chi(z_1)= r_1\, \beta_1$,\ldots, $\chi(z_M) = r_1 \beta_M$ and $\beta_1^2+\ldots+\beta_M^2 = 1$.
The blown up space, ${\cal K}$, is the variety in $\overline{\mathbb{H}}^M\times
\mathbb{R}\times \overline{R} \times\mathbb{R}^{M+1}$ containing all points
$(z_1,\ldots,z_M,q,\lambda,r_1,\beta_1,\ldots,\beta_M)$ that verify the blow-up constraints. The
topology is the one given by the local description as a closed
subset of Euclidean space. The set ${\cal K}\backslash \partial_\infty
{\cal K}$ can be identified with $\mathbb{H}^M\times \mathbb{R}\times \overline{R}$. After
the first blow-up, $\mu_2$ becomes
\begin{eqnarray}
\mu_2&=& \frac{M \prod \limits_{i=1}^M \beta_i
|z_{\lambda}\sum\limits_{i=1}^M z_i+\lambda z_{\lambda}-q
z_{\lambda}+1|^2}{[\sum\limits_{j=1}^M
\prod\limits_{\substack{i=1\\i \neq j}}^M \beta_i]
[\sum\limits_{i=1}^M \beta_i |z_i-z_\lambda |^2 + {\rm
Im}(\lambda)/r_1]}
\end{eqnarray}

We can extend $\mu_2$ to an upper semi-continuous function on
${\cal K}$ by defining, for points $k \in \partial_\infty {\cal K} $,
$$
\mu_2(k) = \limsup_{\substack{k_n \rightarrow k \\ k_n \in
{\cal K}\backslash \partial_\infty {\cal K}}} \mu_2(k_n) \;.
$$
Here $k_n = (z_{1,n},z_{2,n}, \ldots, z_{M,n}, q_{1,n},\lambda_n)$
and it converges to $k$ in ${\cal K}$.

Let us define $\Sigma$ to be the subset of ${\cal K}$ where $\mu_2=1$ and let $\mathcal{K}_0$ denote the subset of $\partial_\infty {\cal K}$ where $\lambda \in (-2\sqrt{M}, 2\sqrt{M})$ and $q=0$. 
For the analysis of $\mu_3$ we need the following lemma:

\begin{lem} \la{L2.3} Let $\Gamma =\{k\in{\cal K}: k=(z_1,\ldots,z_M,0,\lambda,0,\beta,\ldots,\beta)\}\subset {\cal K}$, it contains points in $\mathcal{K}$ with $\beta_1=\ldots=\beta_M=\beta$. Then, $$\Gamma \cap \Sigma \cap \mathcal{K}_0= \{k \in \mathcal{K}_0: k
=(z,\ldots,z,0,\lambda,0,\beta,\ldots,\beta)\}\,.$$
\end{lem}

\proof Let us first derive an upper bound $\mu_2^*$ for $\mu_2$. 
\newline For $k=(z_1,\ldots,z_M,0,\lambda, r_1, \beta_1,\ldots \beta_M)\in {\cal K}\backslash \partial_\infty
{\cal K}$ we have
\begin{align*}
&\mu_2(k) = {{M \,{\rm w}(\phi(z_1,\ldots,z_M,q,\lambda))}\over{\sum\limits_{i=1}^M{\rm w}(z_i)}} = \frac{M \, {\rm c}(z_\lambda,\phi(z_1,\ldots,z_M,q,\lambda))}{\sum\limits_{i=1}^M \mathrm{c}(z_{\lambda}, z_i)} \\
&\hspace{1cm} \leq \frac{M \, {\rm c}(M z_\lambda, z_1
+\ldots+ z_M )}{\sum\limits_{i=1}^Mc(z_\lambda, z_i)} = \frac{M
\,\mathrm{w}(\frac{1}{M}\sum \limits_{i=1}^M z_i)}{\sum\limits_{i=1}^M\mathrm{w}(z_i)}.
\end{align*}
Therefore we can define
\begin{eqnarray}
\mu_2^*(k) =  \frac{M
\,\mathrm{w}(\frac{1}{M}\sum \limits_{i=1}^M z_i)}{\sum\limits_{i=1}^M\mathrm{w}(z_i)} =
\frac{\prod\limits_{i=1}^M \beta_i |\sum\limits_{i=1}^M
(z_i-z_\lambda)|^2}{[\sum\limits_{j=1}^M\prod\limits_{\substack{i=1\\i\neq
j}}^M \beta_i][\sum\limits_{i=1}^M \beta_i|z_i-z_\lambda|^2]} \la{Mu*}
\end{eqnarray}
Clearly $\mu_2 \leq \mu_2^*$, with equality when $\lambda$ is
real.

Let $k\in \Gamma \cap \Sigma \cap {\cal K}_0$. If $k$ is a point of continuity for
$\mu_2^*$ then $\mu_2^*(k)=1$. At a point of continuity $k$, 
$$
1=\mu_2(k)= \limsup_{\substack{k_n\rightarrow k \\ k_n\in
M\backslash
\partial_\infty M}}\mu_2(k_n) \le \limsup_{\substack{k_n\rightarrow
k \\ k_n\in M\backslash \partial_\infty M}}\mu_2^*(k_n) \le 1\,.
$$
The last inequality holds because at a point of continuity, the
$\limsup$ is actually a limit which can be evaluated in any order.
If we take the limit in $\lambda$ and $q$ first, we may use the
fact that for $\lambda\in(-2\sqrt{M},2\sqrt{M})$, $\mu_2 =
\mu_2^*$. Proposition \ref{Prop2} proves that the limit in $z_i$
is at most $1$, which implies $\mu_2^*(k) = 1$ at the points of
continuity.

 Since we do not need to know the entire behavior of
$\mu_2$ at the boundary, we will concentrate only on the situations
needed in the analysis of $\mu_3$. Therefore we need two cases to consider:

\textsc{Case I}: Let $k\in \Gamma\cap \Sigma \cap {\cal K}_0$ such that $z_1,\ldots,z_M \in \partial_\infty \overline{\mathbb{H}}$ and $z_i \neq i \infty$ for all $i=1,\ldots,M$. This is a point of continuity and we have:
$$
\mu_2^*(k)= \frac{ |\sum\limits_{i=1}^M (z_i-z_\lambda)|^2}{M \,\sum\limits_{i=1}^M |z_i-z_\lambda|^2}.
$$
By the triangle inequality and the Cauchy Schwarz inequality,
$$
|\sum\limits_{i=1}^M (z_i-z_\lambda)|^2 \leq
\left[\sum\limits_{i=1}^M |z_i-z_\lambda|\right]^2 \leq M
\left[\sum\limits_{i=1}^M |z_i-z_\lambda|^2\right]\,.
$$
The first inequality turns into equality if $z_i -z_\lambda$ have
the same argument for all $i$ and the second one if $z_i-z_\lambda$
are equal in absolute values. Therefore, $\mu_2^* =1$ iff all $z_i$
are equal.

 \textsc{Case II}: Let $k\in \Gamma \cap \Sigma \cap {\cal K}_0$, $z_1=\ldots=z_a=i\infty$, and $z_{a+1},\ldots,z_{M}$ are real, for some $a$, $1<a<M$. Suppose $(k_n)$
is a sequence that realizes the $\limsup$ in the definition of
$\mu_2(k)$. 
$$
\mu_2^*(k_n)= \frac{ |\sum\limits_{i=1}^a (z_i-z_\lambda)+ \sum\limits_{i=a+1}^M (z_i-z_\lambda)|^2}{M \,\sum\limits_{i=1}^M |z_i-z_\lambda|^2} \leq \frac{ |\sum\limits_{i=1}^a (z_i-z_\lambda)+ \sum\limits_{i=a+1}^M (z_i-z_\lambda)|^2}{M \,\sum\limits_{i=1}^M |z_i-z_\lambda|^2} .
$$
The second term in the numerator stays finite in the limit and therefore, obviously\\ $\dis \mu_2^*(k) \leq \frac{a}{M}$. \qed

We end this section with the proofs of our previous lemmas, Lemma \ref{L2.1} and Lemma \ref{L2.2}.

\noindent \textbf{Proof of Lemma \ref{L2.1}:} In order to simplify the notation, let us define  \\$ Z=(z_1,\ldots, z_{2M-1})$, \,$Q=(q_{1},q_{2})$, $\xi_{\sigma}(Z,Q,\lambda)=(z_{\sigma_1},\dots,z_{\sigma_M},q_{1},\lambda)$,\\ $\tau_{\sigma}(Z,Q,\lambda)=(\phi(\xi_{\sigma}(Z,Q,\lambda)),z_{\sigma_{M+1}},\ldots,z_{\sigma_{2M-1}},q_{2},\lambda)$ and
$$
\nu_i=\frac{\mathrm{w}(z_i)}{\mathrm{w}(z_1)+\ldots+\mathrm{w}(z_{2M-1})}\,.
$$
Extend $\mu_{3,p}$ to an upper semi-continuous function on
$\overline{\mathbb{H}}^{2M-1}\times \mathbb{R}^2 \times \overline{R}$ by setting, at points $Z_0$,
$Q_0$, $\lambda_0$ where it is not already defined,
\begin{eqnarray*}
  \mu_{3,p}(Z_0,Q_0,\lambda_0) &=& \limsup_{Z \rightarrow Z_0, Q\rightarrow Q_0, \lambda\rightarrow \lambda_0}
  \mu_{3,p}(Z,Q,\lambda),\ .
\end{eqnarray*}
The points $Z$, $Q$ and $\lambda$ are approaching their limits in
the topology of $\overline{\mathbb{H}}^{2M-1}\times \mathbb{R}^2 \times \overline{R}$. To prove the
lemma it is enough to show that
\begin{eqnarray*}
  \mu_{3,p}(Z, Q, \lambda) &<& 1
\end{eqnarray*}
for $(Z, Q, \lambda)$ in the compact set
$\partial_\infty\overline{\mathbb{H}}^{2M-1}\times \{0\}^2 \times [-E,E]$, since
this implies that for some $\epsilon > 0$, the upper
semi-continuous function $\mu_{3,p}(Z, Q, \lambda)$ is bounded by
$1-2\epsilon$ on the set, and by $1-\epsilon$ in some
neighborhood. We have

\begin{align*}
&\mu_{3,p}(Z,Q, \lambda) = \sum_{\sigma} {{\mathrm{w}^{1+p}(\phi(\tau_{\sigma}(Z,Q,\lambda)))}\over{ \mathrm{w}^{1+p}(z_1)+\ldots+ \mathrm{w}^{1+p}(z_{2M-1})}}\\
& = \sum_{\sigma} \left({{\mathrm{w}(\phi(\tau_{\sigma}(Z,Q,\lambda)))}\over{ \mathrm{w}(z_1)+\ldots+ \mathrm{w}(z_{2M-1})}}\right)^{1+p} \frac{1}{\nu_1^{1+p}+\ldots+\nu_{2M-1}^{1+p}}=\\
&= \sum_{\sigma}\left[\mu_2(\tau_\sigma)\left(\frac{1}{M^2}\mu_2(\xi_\sigma)(\nu_{\sigma_1}+\ldots+\nu_{\sigma_M})+\frac{1}{M}(\nu_{\sigma_{M+1}}+\ldots+\nu_{\sigma_{2M-1}})
\right) \right]^{1+p} \cdot \\
&\,\,\,\,\,\,\,\,\,\,\,\,\,\,\,\,\,\,\,\,\,\,\,\,\,\,\,\,\,\,\,\,\,\,\,\,\,\,\,\,\,\,\,\,\,\,\,\,\,\,\,\,\,\,\,\,\,\,\,\,\,\,\,\,\,\,\,\,\,\,\,\,\,\,\,\,\,\,\,\,\,\,\,\,\,\,\,\,\,\,\,\,\,\,\,\,\,\,\,\,\,\,\,\,\,\,\,\,\,\,\,\,\,\,\,\,\,\,\,\,\,\,\,\,\,\,\,\,\,\,\,\,\,\,\,\,\,\,\,\,\,\,\,\,\,\,\cdot\frac{1}{\nu_1^{1+p}+\ldots+\nu_{2M-1}^{1+p}} .
\end{align*}
Define $\dis \chi(z_1) = \frac{1}{\mathrm{w}(z_1)} = R_1 \Omega_1$, \ldots, $\dis \chi(z_{2M-1}) = \frac{1}{\mathrm{w}(z_{2M-1})} = R_1 \Omega_{2M-1}$, where $R_1$, $ \Omega_1$, $\Omega_2$,\ldots, $\Omega_{2M-1}$ are defined functions of $Z$ with the property
$\Omega_1^2+\ldots+\Omega_{2M-1}^2=1$. Notice that for any cyclic
permutation $\sigma$,
\begin{eqnarray}
  \nu_{\sigma_l}&=& \frac{\prod\limits_{\substack{j=1\\j \neq l}}^{2M-1} \Omega_{\sigma_j}}{\sum\limits_{i=1}^{2M-1}\left[\prod\limits_{\substack{j=1\\j \neq 1}}^{2M-1}
  \Omega_j\right]} \la{2.13}
\end{eqnarray}

In the analysis of $\mu_2(\xi_\sigma)$ we use the blow-up with
coordinates $r_{1 \sigma}(\xi_\sigma)$ and
$\beta_{\sigma_j}(\xi_\sigma)$ where $j=1,\ldots,M$ and in the
analysis of $\mu_2(\tau_\sigma)$ we use the blow-up with coordinates
$r_{2 \sigma}(\tau_\sigma)$ and $\beta_{\sigma_j}(\tau_\sigma)$
where $j=M,\ldots,2M-1$. Therefore we have the following relations:

$R_1 \Omega_{\sigma_j} = r_{1\sigma} \beta_{\sigma_j}(\xi_\sigma)$
when $j=1,\ldots,M$

$ R_1 F = \chi(\phi(\xi_\sigma)) = r_{2\sigma}
\beta_{\sigma_1}(\tau_\sigma)$

$R_1 \Omega_{\sigma_j} = r_{2\sigma} \beta_{\sigma_j}(\tau_\sigma)$
when $j=M+1,\ldots,2M-1$
\newline where $$ F= \frac{\chi(\phi(\xi_\sigma))}{R_1}= \frac{r_{1\sigma} M \prod\limits_{i=1}^M\beta_{\sigma_i}}{R_1 \mu_2(\xi_\sigma(Z,Q,\lambda))\sum\limits_{j=1}^{M} \prod \limits_{\substack{i=1\\i\neq j}}^M \beta_{\sigma_i}} = \frac{M \Omega_{\sigma_1} \prod\limits_{i=2}^M\beta_{\sigma_i}}{ \mu_2(\xi_\sigma(Z,Q,\lambda))\sum\limits_{j=1}^{M} \prod \limits_{\substack{i=1\\i\neq j}}^M \beta_{\sigma_i}}.
$$
Consequently

$\Omega_{\sigma_j}^2=\beta_{\sigma_j}^2(\xi_\sigma)(\Omega_{\sigma_1}^2+\ldots+\Omega_{\sigma_M}^2)
$ for $j=1,\ldots,M$

$\Omega_{\sigma_j}^2=\beta_{\sigma_j}^2(\tau_\sigma)(F +
\Omega_{\sigma_{M+1}}^2+\ldots+\Omega_{\sigma_{2M-1}}^2) $ for
$j=M,\ldots,2M-1$.

Suppose that $\mu_{3,p}(Z,Q, \lambda)=1$ for some $(Z,Q, \lambda)
\in \partial_\infty\overline{\mathbb{H}}^{2M-1}\times \{0\}^2 \times [-E,E]$. Then
there must exist a sequence $(Z_n,Q_n, \lambda_n)$ with $Z_n 
\longrightarrow Z$ in $\overline{\mathbb{H}}^{2M-1}$, $Q_n \longrightarrow (0,0)$
and $\lambda_n \longrightarrow \lambda \in [-E,E]$ such that
$$
\lim \mu_{3,p}(Z_n,Q_n, \lambda_n)=1.
$$

From now on $Z$ and $\lambda$ will denote the limiting values of
the sequences $Z_n$ and $\lambda_n$. Similarly, we will denote by
$\nu_i$ and $\Omega_i$ the limits of $\nu_i(Z_n)$ and
$\Omega_i(Z_n)$.

We claim that
\begin{equation}
    \nu_1 = \ldots = \nu_{2M-1} = \frac{1}{2M-1}. \la{2.14}
\end{equation}
This follows from the expression for $\mu_{3,p}(Z,Q, \lambda)$,
the bound for $\mu_2$ and the convexity of $x \mapsto x^{1+p}$:
\newline
\begin{align*}
&1 = \mu_{3,p}(Z,Q, \lambda)\\
&=\sum_{\sigma}\left[\mu_2(\tau_\sigma)\left(\frac{1}{M^2}\mu_2(\xi_\sigma)(\nu_{\sigma_1}+\ldots+\nu_{\sigma_M})+\frac{1}{M}(\nu_{\sigma_{M+1}}+\ldots+\nu_{\sigma_{2M-1}})\right) \right]^{1+p} \cdot \\
&\hspace{10cm} \cdot \frac{1}{\nu_1^{1+p}+\ldots+\nu_{2M-1}^{1+p}}\\
&\leq \sum_{\sigma}\left[\left(\frac{1}{M^2}(\nu_{\sigma_1}+\ldots+\nu_{\sigma_M})+\frac{1}{M}(\nu_{\sigma_{M+1}}+\ldots+\nu_{\sigma_{2M-1}})\right)\right]^{1+p} \frac{1}{\nu_1^{1+p}+\ldots+\nu_{2M-1}^{1+p}} \\
& \leq \sum_{\sigma}\left(\frac{1}{M^2}(\nu_{\sigma_1}^{1+p}+\ldots+\nu_{\sigma_M}^{1+p})+\frac{1}{M}(\nu_{\sigma_{M+1}}^{1+p}+\ldots+\nu_{\sigma_{2M-1}}^{1+p})\right) \frac{1}{\nu_1^{1+p}+\ldots+\nu_{2M-1}^{1+p}}= 1,
\end{align*}
so the inequalities must actually be equalities. Since $p>0$,
strict convexity implies that equality only holds if $\nu_1 =
\ldots = \nu_{2M-1}$. Since their sum is $1$, their common value
must be $\dis \frac{1}{2M-1}$.

By going to a subsequence, we may assume that $\Omega_i(Z_n)$
converge. Then (\ref{2.13}) and (\ref{2.14}) imply that their
limiting values along the sequence must be
\begin{equation}
    \Omega_1=\ldots=\Omega_{2M-1}=\frac{1}{\sqrt{2M-1}}.
\end{equation}
One consequence is that
\begin{equation}
z_i \in \partial_{\infty} \overline{\mathbb{H}} \la{2.16}
\end{equation}
for $i=1,\ldots,2M-1$.

Now consider the values of $\xi_\sigma(Z_n,Q_n,\lambda_n)$ and
$\dis \tau_\sigma(Z_n,Q_n,\lambda_n)$. Since these values vary in a
compact region in $\cal M$ we may, again by going to a
subsequence, assume that they converge in $\cal M$ to values which
we will denote $\xi_\sigma$ and $\tau_\sigma$. Using (\ref{2.14})
and the bound $\mu_2 \leq 1$, we find that
\begin{align*}
&1 = \lim_{n \rightarrow \infty} \sum_{\sigma}\left[\mu_2(\tau_\sigma(Z_n,Q_n,\lambda_n))\left(\frac{\mu_2(\xi_\sigma(Z_n,Q_n,\lambda_n))+M-1}{M(2M-1)}
\right) \right]^{1+p} (2M-1)^{p}\\
&\,\,\,\,\,\leq \frac{1}{2M-1} \sum_{\sigma} \left[\frac{1}{M}
\mu_2(\tau_\sigma)\left(\mu_2(\xi_\sigma)+M-1) \right)
\right]^{1+p}\leq 1.
\end{align*}
This implies that for every $\sigma$ occurring in the sum we have $ \dis \mu_2(\xi_\sigma)=\mu_2(\tau_\sigma)=1$.
Therefore, using (\ref{2.16}) we conclude that for each $\sigma$, $\xi_\sigma$ and
$\tau_\sigma$ lie in the set $\Sigma$ given by Lemma \ref{L2.3}.

Now consider the coordinates $\beta_{\sigma_i}$, $i=1,\ldots,M$ for
the point $\xi_\sigma$. These are the limiting values of
$\dis \beta_{\sigma_i}(z_{\sigma_1},\ldots,z_{\sigma_M})$ along our
sequence. Since $\dis 
\Omega_{\sigma_j}^2=\beta_{\sigma_j}^2(\Omega_1^2+\ldots+\Omega_M^2)$
and $\Omega_i=\frac{1}{\sqrt{2M-1}}$, we have $ \beta_{\sigma_i}=\frac{1}{\sqrt{M}}$ for $i=1,\ldots,M$. Going back to the analysis of $\mu_2$, Lemma \ref{L2.3}, we conclude that the
$\overline{\mathbb{H}}$ coordinates of $\xi_\sigma$, namely the limiting values of
$z_{\sigma_1},\ldots,z_{\sigma_M}$ must be equal. Since this is true
for every cyclic permutation, we conclude that
$$
z=z_1=z_2=\ldots=z_{2M-1} \in \partial_\infty \overline{\mathbb{H}}.
$$
We have two distinct cases:

$\bullet$ If $z \in \mathbb{R}$ then
$\phi(z_{\sigma_1},\ldots,z_{\sigma_M},q,\lambda)\longrightarrow
\phi(z,\ldots,z,0,\lambda)= \frac{-1}{Mz+\lambda}$. From the
analysis of $\mu_2$, Case I, the only way
$\tau_\sigma=(\phi(z,\ldots,z,0,\lambda),z,\ldots,z)$ can lie in
$\Sigma$ is if $\phi(z,\ldots,z,0,\lambda)=z$ which would imply
$z=z_\lambda$ and this cannot happen since $z_\lambda \not\in
\partial_\infty \overline{\mathbb{H}}$.

$\bullet$ If $z=i \infty$ then
$\phi(z_{\sigma_1},\ldots,z_{\sigma_M},q,\lambda)\longrightarrow 0$
therefore $\tau_\sigma \longrightarrow (0,i\infty,\ldots,i\infty)$.
Since $\Omega_{\sigma_j}^2=\beta_{\sigma_j}^2(\tau_\sigma)(F +
\Omega_{\sigma_{M+1}}^2+\ldots+\Omega_{\sigma_{2M-1}}^2) $ for
$j=M,\ldots,2M-1$ and $F=\frac{1}{\sqrt{2M-1}}$ in the limiting
case, $\beta_{\sigma_j}(\tau_\sigma)$ are equal. Going back to the
analysis of $\mu_2$, Case II, we conclude that $\mu_2(\tau_\sigma)
<1$.

\noindent Therefore, $ \mu_{3,p}(Z, Q, \lambda) < 1$. \qed
\bigskip

\textbf{Proof of Lemma \ref{L2.2}:} Each term in the sum appearing
in $\mu_{3,p}$ can be estimated

\begin{align*}
&\frac{\mathrm{w} ^{1+p}(\phi(\cdots\cdots))}{\mathrm{w} ^{1+p}(z_1) + \ldots +
\mathrm{w} ^{1+p}(z_{2M-1})} = \frac{(\mathrm{w}(z_1) + \ldots + \mathrm{w}(z_{2M-1}))^{1+p}}{\mathrm{w}^{1+p}(z_1) + \ldots + \mathrm{w}^{1+p}(z_{2M-1})}\cdot \\
& \hspace{9cm}\cdot \left( \frac{\mathrm{w}(\phi(\cdots\cdots))}{\mathrm{w}(z_1) + \ldots +
\mathrm{w}(z_{2M-1})}\right)^{1+p} \\
&\hspace{5cm} \le (2M-1)^{p}\left(\frac{\mathrm{w}(\phi(\cdots\cdots))}{\mathrm{w}(z_1) + \ldots + \mathrm{w}(z_{2M-1})}\right)^{1+p} \;,
\end{align*}
where $\phi(\cdots\cdots)$ denotes $\phi(\phi(z_{\sigma_1},\ldots,
z_{\sigma_M},q_{\sigma_1},\lambda), z_{\sigma_{M+1}}, \ldots,
z_{\sigma_{2M-1}},q_{\sigma_2},\lambda)$. Therefore it is enough to
prove
$$
\frac{\mathrm{w}(\phi(\cdots\cdots))}{\mathrm{w}(z_1) + \ldots + {\rm
cd}(z_{2M-1})} \le C(1+\sum_{i=1}^2 |q_i|^2)\,.
$$
Let $\phi(\cdots)$ denote $\phi(z_{\sigma_1}, \ldots, z_{\sigma_M},
q_1, \lambda)$. We have
\begin{align*}
&\frac{ \mathrm{w}(\phi(\cdots\cdots))}{ \mathrm{w}(z_1) + \ldots + {\rm \mathrm{w}}(z_{2M-1})}= \frac{\left|1+ z_\lambda
\left(\phi(\ldots)+\sum\limits_{i=M+1}^{2M-1}z_{\sigma_i}+\lambda-q_{\sigma_2} \right)  \right|^2}{{\rm Im}\left[\phi(\ldots)+\sum\limits_{i=M+1}^{2M-1}z_{\sigma_i}+\lambda \right ]} \cdot
\frac{1}{\sum\limits_{i=1}^{2M-1}\frac{|z_i-z_\lambda|^2}{{\rm Im}(z_i)}} = \\ 
&= \frac{|\sum\limits_{i=1}^{M}z_{\sigma_i} +\lambda-q_{\sigma_1} + z_\lambda (-1+(\sum\limits_{i=1}^{M}z_{\sigma_i}+\lambda -q_{\sigma_1})(\sum\limits_{i=M+1}^{2M-1}z_{\sigma_i}+\lambda -q_{\sigma_2}))|^2}{{\rm Im}(\sum\limits_{i=1}^{M}z_{\sigma_i}+\lambda)+{\rm Im}(\sum\limits_{i=M+1}^{2M-1}z_{\sigma_i}+\lambda)|\sum\limits_{i=1}^{M}z_{\sigma_i}+\lambda-q_{\sigma_1}|^2} \cdot \frac{1}{\sum\limits_{i=1}^{2M-1}\frac{|z_i-z_\lambda|^2}{{\rm Im}(z_i)}}\\ 
&\leq C \left(\frac{1}{{\rm Im}(\sum\limits_{i=M+1}^{2M-1}z_{\sigma_i})} +\frac{|-1 + (\sum\limits_{i=M+1}^{2M-1}z_{\sigma_i} +\lambda-q_{\sigma_2})(\sum\limits_{i=1}^{M}z_{\sigma_i} +\lambda-q_{\sigma_1})|^2}{{\rm Im}(\sum\limits_{i=1}^{M}z_{\sigma_i})+{\rm Im}(\sum\limits_{i=M+1}^{2M-1}z_{\sigma_i})|\sum\limits_{i=1}^{M}z_{\sigma_i}+\lambda -q_{\sigma_1} |^2}\right)\cdot
\frac{1}{\sum\limits_{i=1}^{2M-1}\frac{|z_i-z_\lambda|^2}{{\rm Im}(z_i)}} 
\end{align*}
\newpage
\begin{align*}
&\leq C\left(\frac{1}{{\rm Im}(\sum\limits_{i=M+1}^{2M-1}z_{\sigma_i})} +2\left( \frac{1}{{\rm Im}(\sum\limits_{i=1}^{M}z_{\sigma_i})}+\frac{|\sum\limits_{i=M+1}^{2M-1}z_{\sigma_i}+\lambda -q_{\sigma_2}|^2|\sum\limits_{i=1}^{M}z_{\sigma_i} +\lambda-q_{\sigma_1}|^2 }{{\rm Im}(\sum\limits_{i=1}^{M}z_{\sigma_i})+{\rm Im}(\sum\limits_{i=M+1}^{2M-1}z_{\sigma_i})|\sum\limits_{i=1}^{M}z_{\sigma_i}+\lambda -q_{\sigma_1} |^2}\right)\right)\cdot \\
&\hspace{11.8cm} \cdot \frac{1}{\sum\limits_{i=1}^{2M-1}\frac{|z_i-z_\lambda|^2}{{\rm Im}(z_i)}} \\ 
&\leq C \left(\frac{1}{{\rm Im}(\sum\limits_{i=M+1}^{2M-1}z_{\sigma_i})} +2\left( \frac{1}{{\rm Im}(\sum\limits_{i=1}^{M}z_{\sigma_i})}+\frac{|\sum\limits_{i=M+1}^{2M-1}z_{\sigma_i}+\lambda -q_{\sigma_2}|^2}{{\rm Im}(\sum\limits_{i=M+1}^{2M-1}z_{\sigma_i})}\right)\right)\cdot \frac{1}{\sum\limits_{i=1}^{2M-1}\frac{|z_i-z_\lambda|^2}{{\rm Im}(z_i)}}\,.
\end{align*}

Choose the compact set $\cal M$ so that $\dis \sum\limits_{i=1}^{2M-1} |z_i-z_\lambda|^2 / {\rm Im}(z_i)\ge C >0$ for some constant $C$ and $(z_1,\ldots,z_{2M-1})\in {\cal M}^c$. Then we can estimate each term depending on whether $z_{\sigma_i}$ is close to $z_\lambda$. 

If all $z_{\sigma_i}$ are sufficiently close to $z_\lambda$, then ${\rm Im}(z_{\sigma_i})$ is bounded below and $|z_{\sigma_i}|$ is bounded above by a constant. Thus
$$
{\rm Im}\left(\sum\limits_{i=M+1}^{2M-1}z_{\sigma_i}\right)\sum_{i=1}^{2M-1}
|z_i-z_\lambda|^2 / {\rm Im}(z_i)\ge {\rm
Im}\left(\sum\limits_{i=M+1}^{2M-1}z_{\sigma_i}\right)C \ge C'
> 0\,,
$$
$$
{\rm Im}\left(\sum\limits_{i=1}^{M}z_{\sigma_i}\right)\sum_{i=1}^{2M-1}
|z_i-z_\lambda|^2 / {\rm Im}(z_i)\ge {\rm Im}\left(\sum\limits_{i=1}^{M}z_{\sigma_i}\right)C \ge C'
> 0
$$
and
\begin{align*}
&\Big|\sum\limits_{i=M+1}^{2M-1}z_{\sigma_i}+\lambda-q_{\sigma_2}\Big|^2
\le\left(\Big|\sum\limits_{i=M+1}^{2M-1}z_{\sigma_i}+\lambda\Big|+|q_{\sigma_2}|\right)^2
\le\left(\Big|\sum\limits_{i=M+1}^{2M-1}z_{\sigma_i}+\lambda\Big|^2+1\right)\left(|q_{\sigma_2}|^2+1\right)\\
&\le\left(\left(\Big|\sum\limits_{i=M+1}^{2M-1}z_{\sigma_i}\Big|+|\lambda|\right)^2+1\right)\left(|q_{\sigma_2}|^2+1\right) \le \left(\Big|\sum\limits_{i=M+1}^{2M-1}z_{\sigma_i}\Big|^2\left(|\lambda|^2+1\right)+1\right)\left(|q_{\sigma_2}|^2+1\right)
\end{align*}
$\dis \le C \left(\Big|\sum\limits_{i=M+1}^{2M-1}z_{\sigma_i}\Big|^2+1\right)\left(|q_{\sigma_2}|^2+1\right) \le C\left(1 + |q_{\sigma_2}|^2\right)$, so we are done. 

If all $z_{\sigma_i}$ are far from $z_\lambda$,
$\dis {\rm Im}(\sum\limits_{i=M+1}^{2M-1}z_{\sigma_i})\sum_{i=1}^{2M-1}
|z_i-z_\lambda|^2 / {\rm Im}(z_i) \ge \sum\limits_{i=M+1}^{2M-1}|z_{\sigma_i}-z_\lambda|^2 \ge$\\$\dis 
\frac{1}{M-2}|\sum\limits_{i=M+1}^{2M-1}(z_{\sigma_i}-z_\lambda)|^2
\ge$$\dis C(1+|\sum\limits_{i=M+1}^{2M-1}z_{\sigma_i}|^2)$
so that\\
$\dis |\sum\limits_{i=M+1}^{2M-1}z_{\sigma_i}+\lambda-q_{\sigma_2}|^2 \Big/
\left({\rm
Im}(\sum\limits_{i=M+1}^{2M-1}z_{\sigma_i})\sum\limits_{i=1}^{2M-1}
|z_i-z_\lambda|^2/ {\rm Im}(z_i)\right) \le C(1 + |q_{\sigma_2}|^2)$
in this case too. Also,
$$
{\rm Im}(\sum\limits_{i=1}^{M}z_{\sigma_i})\sum_{i=1}^{2M-1}
|z_i-z_\lambda|^2 / {\rm Im}(z_i) \ge
\sum\limits_{i=1}^{M}|z_{\sigma_i}-z_\lambda|^2 \ge
C(1+|\sum\limits_{i=1}^{M}z_{\sigma_i}|^2)\,.
$$

If at least one $z_{\sigma_j}$ is not close to $z_\lambda$ for
$j=1,\ldots, M$, the first term is still bounded. If at least one
$z_{\sigma_j}$ is close to $z_\lambda$ for $j=M+1, \ldots, 2M-1$,
then the second term is finite and
$$
{\rm Im}(\sum\limits_{i=M+1}^{2M-1}z_{\sigma_i})\sum_{i=1}^{2M-1}
|z_i-z_\lambda|^2 / {\rm Im}(z_i) \ge C + |z_{\sigma_j}-z_\lambda|^2
\ge C(C+|z_{\sigma_j}|^2)\,.
$$
Therefore
\begin{align*}
\frac{\Big|\sum\limits_{i=M+1}^{2M-1}z_{\sigma_i}+\lambda-q_{\sigma_2}\Big|^2}{
{\rm
Im}\left(\sum\limits_{i=M+1}^{2M-1}z_{\sigma_i}\right)\sum\limits_{i=1}^{2M-1}
|z_i-z_\lambda|^2/ {\rm Im}(z_i)} \le C \frac{\left(C_1 +
|z_{\sigma_j}|^2\right)\left(1 + |q_{\sigma_2}|^2\right)}{ C_2 + |z_{\sigma_j}|^2}\le
C\left(1 + |q_{\sigma_2}|^2\right).
\end{align*}

The estimates for $\mu'_{3,p}$ are very similar. We omit the details.
 \qed

\end{document}